\documentclass[aps,superscriptaddress,prc,twocolumn,nofootinbib]{revtex4}

\usepackage{graphicx}
\usepackage{epstopdf}
\usepackage{subfigure}
\usepackage{epsfig}
\usepackage{amsmath,amssymb,amsfonts}
\usepackage{color}
\usepackage[utf8]{inputenc}

\begin{document}

\title{Linearized Boltzmann transport model for jet propagation in the quark-gluon plasma: Heavy quark evolution}

\author{Shanshan Cao}
\affiliation{Nuclear Science Division, Lawrence Berkeley National Laboratory, Berkeley, CA 94720, USA}
\author{Tan Luo}
\affiliation{Institute of Particle Physics and Key Laboratory of Quark and Lepton Physics (MOE), Central China Normal University, Wuhan, 430079, China}
\affiliation{Nuclear Science Division, Lawrence Berkeley National Laboratory, Berkeley, CA 94720, USA}
\author{Guang-You Qin}
\affiliation{Institute of Particle Physics and Key Laboratory of Quark and Lepton Physics (MOE), Central China Normal University, Wuhan, 430079, China}
\author{Xin-Nian Wang}
\affiliation{Institute of Particle Physics and Key Laboratory of Quark and Lepton Physics (MOE), Central China Normal University, Wuhan, 430079, China}
\affiliation{Nuclear Science Division, Lawrence Berkeley National Laboratory, Berkeley, CA 94720, USA}

\date{\today}


\begin{abstract}

A Linearized Boltzmann Transport (LBT) model coupled with hydrodynamical background is established to describe the evolution of jet shower partons and medium excitations in high energy heavy-ion collisions. We extend the LBT model to include both elastic and inelastic processes for light and heavy partons in the quark-gluon plasma. A hybrid model of fragmentation and coalescence is developed for the hadronization of heavy quarks. Within this framework, we investigate how heavy flavor observables depend on various ingredients, such as different energy loss and hadronization mechanisms, the momentum and temperature dependences of the transport coefficients, and the radial flow of the expanding fireball. Our model calculations show good descriptions of the $D$ meson suppression and elliptic flow observed at the LHC and RHIC. The prediction for the Pb-Pb collisions at $\sqrt{s_\mathrm{NN}}$=5.02~TeV is provided.

\end{abstract}

\maketitle


\section{Introduction}
\label{sec:Introduction}

The nuclear suppression of high transverse momentum ($p_\mathrm{T}$) hadrons and their azimuthal anisotropy observed at the Relativistic Heavy-Ion Collider (RHIC) \cite{Adams:2003kv,Abelev:2008ae,Adare:2012wg,Adare:2014bga} and the Large Hadron Collider (LHC) \cite{Aamodt:2010jd,Abelev:2012di,CMS:2012aa,Chatrchyan:2012xq,ATLAS:2011ah} provide valuable evidence of the formation of the strongly interacting quark-gluon plasma (sQGP) in high energy nucleus-nucleus collisions. 
Such phenomena are commonly understood as the consequence of energy loss experienced by high energy partons produced via initial hard processes as they propagate through and interact with the color-deconfined QGP before fragmenting into hadrons \cite{Qin:2015srf, Wang:1991xy, Gyulassy:1993hr, Baier:1996kr, Zakharov:1996fv}. 
Parton energy loss can be attributed to both elastic (collisional) \cite{Braaten:1991we, Qin:2007rn} and inelastic (radiative) \cite{Wang:1991xy, Gyulassy:1993hr} parton-medium interaction processes. 
Phenomenological studies of parton energy loss involve sophisticated calculations of the quenching of single inclusive hadron \cite{Bass:2008rv, Armesto:2009zi, Chen:2010te}, di-hadron \cite{Zhang:2007ja,Renk:2008xq, Majumder:2004pt}, photon-hadron correlations \cite{Renk:2006qg,Zhang:2009rn,Qin:2009bk,Wang:2013cia}, as well as full jets \cite{Qin:2010mn,Dai:2012am,Wang:2013cia,Chang:2016gjp}. 
Various approaches to parton energy loss have been utilized in these phenomenological applications, and one may refer to Ref. \cite{Armesto:2011ht} for a detailed comparison between different schemes. 
One of the major goals of these studies is to extract quantitatively jet transport coefficients in the QGP from systematic comparison with experimental data \cite{Burke:2013yra}.

Among various hadron species, heavy flavor hadrons are of great interest. 
High $p_\mathrm{T}$ heavy quarks provide a valuable tool to study the energy loss of energetic partons. Heavy quarks with low $p_\mathrm{T}$ may be thermalized with the medium background within the life-time of the QGP fireball and therefore can encode rich information of the thermal properties of the medium. In the intermediate $p_\mathrm{T}$ region, heavy flavor hadrons provide a unique opportunity for studying the hadronization mechanism of energetic partons. 
Over the past decade, experiments at both RHIC and the LHC have produced many interesting measurements for heavy flavor hadrons and their decay electrons. 
The most surprising observations are the small values of the nuclear modification factors $R_\mathrm{AA}$ and the large values of the elliptic flow coefficients $v_2$; both are almost comparable to those of light hadrons \cite{Adare:2010de,Adare:2014rly,Adamczyk:2014uip,Xie:2016iwq,Lomnitz:2016rpz,ALICE:2012ab,Abelev:2013lca}.
These results seem contradictory to the earlier expectation of the mass hierarchy in parton energy loss \cite{Dokshitzer:2001zm}.
Thus it still remains a challenge to obtain a full understanding of the dynamical evolution of different flavor partons in heavy-ion collisions, including their initial production, in-medium energy loss and the hadronization process.

Various transport models have been constructed to study the heavy flavor production and medium modification in heavy-ion collisions, such as the parton cascade models based on the Boltzmann equation \cite{Molnar:2006ci,Zhang:2005ni,Uphoff:2011ad,Uphoff:2012gb}, the linearized Boltzmann \cite{Gossiaux:2010yx,Nahrgang:2013saa} and Langevin \cite{Svetitsky:1987gq,Moore:2004tg, Akamatsu:2008ge,Das:2010tj,He:2011qa,Young:2011ug,Alberico:2011zy,Cao:2013ita,Cao:2015hia} approaches coupled to hydrodynamic background, and the parton-hadron-string-dynamics model \cite{Song:2015ykw}. 
In this work, we extend the Linearized Boltzmann Transport (LBT) model \cite{Wang:2013cia,He:2015pra} to include both light and heavy flavor parton evolution in relativistic heavy-ion collisions. 
All $2\rightarrow 2$ channels are included for the elastic scattering processes, and the medium-induced gluon radiation is introduced for the inelastic process based on the higher-twist energy loss formalism \cite{Guo:2000nz,Majumder:2009ge,Zhang:2003wk}.
Within this framework, we obtain good descriptions of the $D$ meson suppression and elliptic flow coefficient observed at RHIC and the LHC.
We also provide the prediction for Pb-Pb collisions at $\sqrt{s_\mathrm{NN}}$=5.02~TeV. 
In this work, we focus on heavy flavor observables; the calculations for light flavor hadrons and jets will be presented separately in our upcoming studies.

The paper is organized as follows. 
In Sec. \ref{sec:LBT}, we provide a brief overview of the LBT model and present the extension for heavy quark evolution in dense nuclear medium. Both collisional and radiative processes experienced by heavy quarks will be studied inside a static thermal medium. 
In this section, we also validate our model by comparing with semi-analytical calculations.
In Sec. \ref{sec:hadronization}, we develop a hybrid model of fragmentation and coalescence to describe the hadronization process of heavy quarks into heavy flavor hadrons.
In Sec. \ref{sec:results}, we couple our LBT model with hydrodynamic background, and calculate heavy meson suppression and elliptic flow coefficient for heavy-ion collisions at the LHC and RHIC. 
The effects of different ingredients in the transport model on the final heavy meson suppression and elliptic flow will be investigated in detail, such as different momentum and temperature dependences of the transport coefficients, different hadronization mechanisms, and the radial flow of the hydrodynamically expanding medium. 
Sec. \ref{sec:summary} contains the summary and the outlook for future developments.

\section{Heavy quark evolution within a linearized Boltzmann transport model}
\label{sec:LBT}

\subsection{Elastic scattering}
\label{subsec:collision}

The evolution of the phase space distribution of a given parton (noted as ``1") in the absence of a mean field can be described using the Boltzmann equation:
\begin{eqnarray}
  \label{eq:boltzmann1}
  p_1\cdot\partial f_1(x_1,p_1)=E_1 C\left[f_1\right],
\end{eqnarray}
where  $p_1=(E_1,\vec{p}_1)$ is the 4-momentum of the jet parton ``1" and $C[f_1]$ denotes the collision integral which can
be written as
\begin{eqnarray}
  \label{eq:collision}
  C\left[f_1\right]\equiv 
   \int d^3 k \!\!&&\!\! \left[w(\vec{p}_1+\vec{k},\vec{k})f_1(\vec{p}_1+\vec{k}) \right. \nonumber\\
  &&\left. -w(\vec{p}_1,\vec{k})f_1(\vec{p}_1)\right] , 
\end{eqnarray}
where $w(\vec{p}_1,\vec{k})$ denotes the transition rate for parton ``1" from the momentum state $\vec{p}_1$ to $\vec{p}_1-\vec{k}$, 
which can be directly related to the microscopic scattering cross section of  parton ``1". For elastic scattering processes ``$1+2\rightarrow3+4$",
\begin{align}
 w(\vec{p}_1,\vec{k}) & \equiv \sum_{2,3,4} w_{12\rightarrow 34}(\vec{p}_1,\vec{k}),\\
  \label{eq:transition}
  w_{12\rightarrow 34}(&\vec{p}_1,\vec{k})= \gamma_2\int\frac{d^3 p_2}{(2\pi)^3}f_2(\vec{p}_2)\left[1\pm f_3(\vec{p}_1-\vec{k}) \right]\nonumber\\
&\times\left[1\pm f_4(\vec{p}_2+\vec{k})\right] v_{\textnormal{rel}} \nonumber\\
&\times d\sigma_{12\rightarrow 34}(\vec{p}_1,\vec{p}_2\rightarrow \vec{p}_1-\vec{k},\vec{p}_2+\vec{k}),
\end{align}
where the summation runs over all flavors in all possible ``$1+2\rightarrow3+4$" channels, $\gamma_2$ represents the spin-color degeneracy of parton ``2" (6 for a quark and 16 for a gluon), and $v_{\textnormal{rel}}=\sqrt{(p_1\cdot p_2)^2-(m_1 m_2)^2}/(E_1 E_2)$ is the relative velocity between ``1" and ``2". 
The ``$\pm$" signs in Eq. (\ref{eq:transition}) denote the Bose-enhancement or Pauli-blocking effect. 

The elastic scattering rate for  parton ``1" through a given channel can be obtained by integrating the transition rate over the exchange momentum $\vec{k}$:
\begin{eqnarray}
 \label{eq:rate0}
 \Gamma_\mathrm{12\rightarrow34}(\vec{p}_1)=\int d^3k w_{12\rightarrow 34}(\vec{p}_1,\vec{k}).
\end{eqnarray}
Using the expression for $v_\mathrm{rel}d\sigma_{12\rightarrow 34}$, one obtains:
\begin{align}
 \label{eq:rate1}
 \Gamma_\mathrm{12\rightarrow34}&(\vec{p}_1) = \frac{\gamma_2}{2E_1}\int \frac{d^3 p_2}{(2\pi)^3 2E_2}\int\frac{d^3 p_3}{(2\pi)^3 2E_3}\int\frac{d^3 p_4}{(2\pi)^3 2E_4}\nonumber\\
&\times f_2(\vec{p}_2)\left[1\pm f_3(\vec{p}_3) \right]\left[1\pm f_4(\vec{p}_4)\right]S_2(s,t,u)\nonumber\\
&\times (2\pi)^4\delta^{(4)}(p_1+p_2-p_3-p_4)|\mathcal{M_\mathrm{12\rightarrow34}}|^2,
\end{align}
where the factor
\begin{eqnarray}
 \label{eq:S2}
 S_2(s,t,u)=\theta(s\ge2\mu_\mathrm{D}^2)\theta(-s+\mu_\mathrm{D}^2\le t\le -\mu_\mathrm{D}^2)
\end{eqnarray}
is introduced \cite{He:2015pra,Auvinen:2009qm} to avoid possible divergence of $|\mathcal{M_\mathrm{12\rightarrow34}}|^2$ at small angle ($u,t\rightarrow 0$) for massless partons, and $\mu_\mathrm{D}^2=g^2T^2(N_c+N_f/2)/3$ is the Debye screening mass in the QGP medium. 
In our model, the leading-order matrix elements are taken for elastic scattering processes (see Ref. \cite{Eichten:1984eu} for massless light partons and Ref. \cite{Combridge:1978kx} for heavy quarks). 

In this paper, we focus on the evolution of heavy quarks, i.e., partons ``1" and ``3" represent the incoming and outgoing heavy quarks, ``2" represents a light parton from the QGP background and ``4" represents the corresponding recoiled thermal parton after scattering.  Therefore in Eq. (\ref{eq:rate1}), $f_2$ and $f_4$ are taken as thermal distributions for massless partons; the factor $(1-f_3)$ for heavy quarks can be neglected due to their dilute density inside the QGP when the temperature is much smaller than the heavy quark mass.

With the above setup, we obtain the following elastic scattering rate for a heavy quark inside a thermal medium:
\begin{align}
 \label{eq:rate2}
 \Gamma_\mathrm{12\rightarrow34}&(\vec{p}_1,T) = \frac{\gamma_2}{16E_1(2\pi)^4}\int dE_2 d\theta_2 d\theta_4 d\phi_{4}\nonumber\\
&\times f_2(E_2,T)\left[1\pm f_4(E_4,T)\right] S_2(s,t,u)|\mathcal{M_\mathrm{12\rightarrow34}}|^2\nonumber\\
&\times \frac{E_2 E_4 \sin \theta_2 \sin \theta_4}{E_1-|\vec{p}_1| \cos\theta_{4}+E_2-E_2\cos\theta_{24}},
\end{align}
where
\begin{align}
 \label{eq:E4}
 \cos\theta_{24}&=\sin\theta_2\sin\theta_4\cos\phi_{4}+\cos\theta_2\cos\theta_4,\\[10pt]
 E_4=&\frac{E_1E_2-p_1E_2\cos\theta_{2}}{E_1-p_1\cos\theta_{4}+E_2-E_2\cos\theta_{24}}.
\end{align}
In the above expressions, we choose $\vec{p}_1$ in the $+z$ direction and $\vec{p}_2$ in the $x-z$ plane. $\theta_2$ is the polar angle of $\vec{p}_2$, $\theta_4$ and $\phi_4$ are the polar and azimuthal angles of $\vec{p}_4$, and $\theta_{24}$ is the angle between $\vec{p}_2$ and $\vec{p}_4$.

In our model, given the momentum of a heavy quark and the local temperature of its surrounding medium, one may first calculate the scattering rate $\Gamma_i$ for each channel $i=(12\rightarrow 34)$ (with gluon or light quarks). The total rate is the sum of all channels: $\Gamma=\sum_i\Gamma_i$.
The scattering probability during a time interval $\Delta t$ is given by: $P_\mathrm{el}=\Gamma\Delta t$.\footnote{For sufficiently small $\Delta t$, e.g. 0.1~fm, $P_\mathrm{el}\ll1$ can be satisfied in realistic QGP medium and thus can be interpreted as the probability.}
With the total scattering probability, one may determine whether or not the heavy quark is scattered for a given time interval $\Delta t$. 
If it is scattered, a specific channel is then selected according to its relative probability $\Gamma_i/\Gamma$. 
The momenta of the outgoing heavy quark and recoiled light parton are sampled according to the differential interaction rate [Eq. (\ref{eq:rate2})].

For convenience, we use Eq. (\ref{eq:rate2}) to define the following notation: 
\begin{align}
 \label{eq:defX}
 \langle\langle &X(\vec{p}_1,T) \rangle\rangle = \sum_\mathrm{12\rightarrow34}\frac{\gamma_2}{16E_1(2\pi)^4}\int dE_2 d\theta_2 d\theta_4 d\phi_{4}\nonumber\\
&\times X(\vec{p}_1,T) f_2(E_2,T)\left[1\pm f_4(E_4,T)\right] S_2(s,t,u)\nonumber\\
&\times|\mathcal{M_\mathrm{12\rightarrow34}}|^2\frac{E_2 E_4 \sin \theta_2 \sin \theta_4}{E_1-|\vec{p}_1| \cos\theta_{4}+E_2-E_2\cos\theta_{24}}.
\end{align}
In particular, we have $\Gamma = \langle\langle 1 \rangle\rangle$ and
\begin{eqnarray}
 \label{eq:22coefficients}
\hat{q}=\langle\langle (\vec{p}_3 - \hat{p}_1\cdot \vec{p}_3)^2\rangle\rangle,\quad \hat{e}=\langle\langle (E_1-E_3)\rangle\rangle,
\end{eqnarray}
where $\hat{q}$ and $\hat{e}$ denote the momentum broadening and energy loss of heavy quark, respectively,  per unit time due to elastic scattering.

\begin{figure}[tb]
  \epsfig{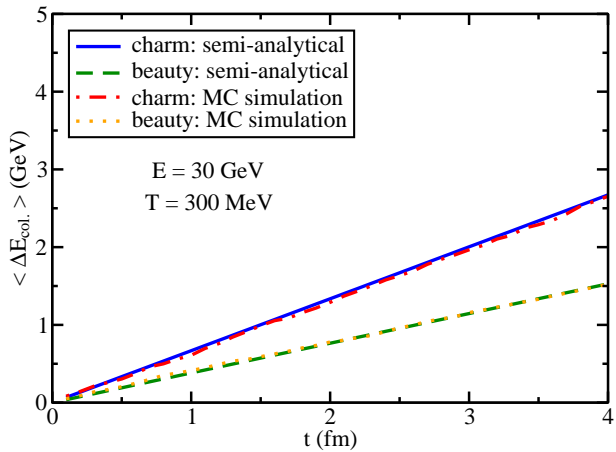}
  \caption{(Color online) Heavy quark energy loss due to elastic scatterings in a static medium: semi-analytical calculation vs. Monte-Carlo simulation.}
 \label{fig:plot-tHQeC}
\end{figure}

In Fig. \ref{fig:plot-tHQeC}, we present the cumulative elastic energy loss of 30~GeV charm and beauty quarks in a static medium with 300~MeV temperature as functions of time. Our LBT Monte Carlo simulation is validated by being compared to the semi-analytical calculation. Here we use a fixed strong coupling constant $\alpha_s=0.3$. 
We observe that our Monte-Carlo simulation of the elastic scattering processes produces an elastic  energy loss that increases linearly with time with a slope of $\hat{e}=0.668$~GeV/fm for charm quarks and 0.382~GeV/fm for beauty quarks. These agree with the values from semi-analytical calculations using Eq. (\ref{eq:22coefficients}). 
Note that in order to consistently compare to the semi-analytical results, we restore the initial energy for heavy quarks after each time step, since the semi-analytical calculation does not automatically include the momentum variation of heavy quarks with time during the propagation.

\subsection{Medium-induced gluon radiation}
\label{subsec:radiation}

When energetic partons propagate through the dense nuclear medium, they may also lose energy via medium-induced gluon radiation. We include the inelastic scattering process in our current LBT model by following our earlier work \cite{Wang:2013cia,Cao:2015hia,Cao:2013ita}.

Given a time step $\Delta t$ at time $t$, the average number of radiated gluons from a hard parton with energy $E$ inside a thermal medium of temperature $T$ can be obtained as:
\begin{equation}
 \label{eq:gluonnumber}
 \langle N_g \rangle(E,T,t,\Delta t) = \Delta t \int dxdk_\perp^2 \frac{dN_g}{dx dk_\perp^2 dt},
\end{equation}
where the gluon radiation spectrum is adopted from the higher-twist calculation for parton energy loss from medium-induced radiation process. 
The distribution function of radiated gluons emitted from a massless parton is calculated in Refs. \cite{Guo:2000nz,Majumder:2009ge} and its modification due to the mass effect of heavy quarks is investigated in Ref. \cite{Zhang:2003wk}:
\begin{align}
\label{eq:gluondistribution}
\frac{dN_g}{dx dk_\perp^2 dt}=\frac{2\alpha_s C_A P(x)}{\pi k_\perp^4} \hat{q}\left(\frac{k_\perp^2}{k_\perp^2+x^2 M^2}\right)^4{\sin}^2\left(\frac{t-t_i}{2\tau_f}\right).
\end{align}
Here, $x$ is the fractional energy of the emitted gluon taken from its parent parton, and $k_\perp$ is the gluon transverse momentum. $\alpha_s$ is the strong coupling constant, $\hat{q}$ is the quark transport coefficient due to elastic scatterings as given by Eq. (\ref{eq:22coefficients}), and $\tau_f$ is the formation time of the radiated gluon defined as $\tau_f={2Ex(1-x)}/{(k_\perp^2+x^2M^2)}$ with $M$ being the mass of the parent parton. $t_i$ denotes the ``initial time", i.e., the production time of the ``parent" parton from which the gluon is then radiated. 
In our calculation, we use the following splitting functions for quarks and gluons respectively \cite{Wang:2009qb}:
\begin{align}
 \label{eq:splitting}
 &P_{q\rightarrow qg}=\frac{(1-x)(2-2x+x^2)}{x},\\
 &P_{g\rightarrow gg}=\frac{2(1-x+x^2)^3}{x(1-x)}.
\end{align}
Note that the color factor has been factorized in the gluon radiation spectrum in Eq. (\ref{eq:gluondistribution}).

When calculating the average gluon number, a lower cut-off of radiated gluon energy $\omega_0=\mu_D$ is imposed to avoid the divergence at $x\rightarrow 0$. Since $\langle N_g\rangle$ may not always be much smaller than 1 during the parton evolution, we allow multiple gluon radiation during each time step. For the $q\rightarrow qg$ process, the following Poisson distribution is used for the number $n$ of radiated gluons during $\Delta t$:
\begin{align}
 \label{eq:Poisson}
 P(n)=\frac{\langle N_g \rangle^n}{n!}e^{-\langle N_g \rangle},
\end{align}
from which one may obtain the total probability of inelastic scattering, i.e, the probability of radiating as least one gluon, as follows:
\begin{align}
 \label{eq:Pinel}
P_\mathrm{inel}=1-e^{-\langle N_g \rangle}.
\end{align}
During each time step in our simulations, the average number of radiated gluons $\langle N_g \rangle$ from a propagating quark is first evaluated using Eq. (\ref{eq:gluonnumber}). 
Then Eq. (\ref{eq:Pinel}) is used to determine whether or not the quark radiates gluons. 
If the gluon radiation process happens, the number $n$ of radiated gluons is sampled following Eq. (\ref{eq:Poisson}), and the energy and momentum of each gluon is sampled according to its spectrum Eq. (\ref{eq:gluondistribution}). 
For the $g\rightarrow gg$ process, $\langle N_g\rangle/2$ is used instead of $\langle N_g\rangle$ in Eq. (\ref{eq:Poisson}) and (\ref{eq:Pinel}), since unlike the $q\rightarrow qg$ process, the number of splittings here is half of the final gluon number.
Note that in our framework, the gluon radiation process is induced by the scattering between the incoming hard parton and the background medium, i.e., a $2\rightarrow 2$ scattering is first assumed for the gluon radiation process. 
Therefore, the four-momenta of the two outgoing partons as presented in Sec. \ref{subsec:collision} need to be adjusted together with the $n$ radiated gluons so that the energy-momentum conservation of the $2\rightarrow 2+n$ inelastic process is respected.

\begin{figure}[tb]
  \epsfig{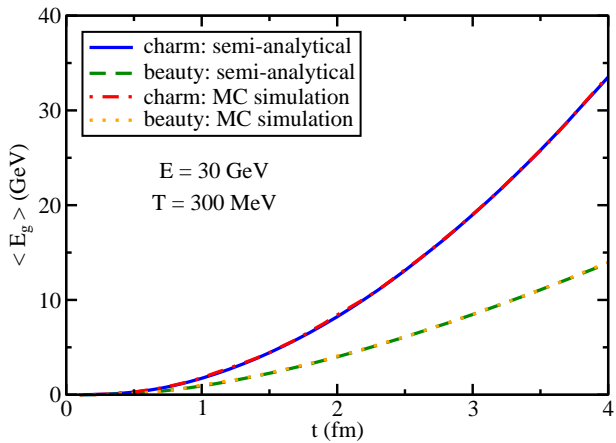}
  \caption{(Color online) Energy of radiated gluons from heavy quarks in a static medium: semi-analytical calculation vs. Monte-Carlo simulation.}
 \label{fig:plot-tHQeG}
\end{figure}

To verify the accuracy of our model simulation, we calculate the cumulative energy carried away by radiated gluons from heavy quarks as functions of time in Fig. \ref{fig:plot-tHQeG}, compared to the results from the following semi-analytical evaluation:
\begin{equation}
 \label{eq:gluonenergy}
\langle E_g\rangle(E,T,t) = \int_0^t dt\int_{\frac{\omega_0}{E}}^1dx\int_0^{(xE)^2}dk_\perp^2 xE\frac{dN_g}{dx dk_\perp^2 dt}.\nonumber
\end{equation}
With $\alpha_s=0.3$, for 30~GeV energy of heavy quarks and 300~MeV temperature of the medium, we have $\hat{q}/T^3$ = 6.25 for charm quarks and 5.24 for beauty quarks as given by Eq. (\ref{eq:22coefficients}). For an apple-to-apple comparison to the semi-analytical calculation, the heavy quark energy is restored to 30~GeV after each inelastic scattering in our simulation and the initial time $t_i$ is fixed at $0$. With these setups, our Monte-Carlo simulation provides a good agreement with the semi-analytical calculation.

\subsection{Elastic versus inelastic energy loss}
\label{subsec:colVSrad}

\begin{figure}[tb]
  \epsfig{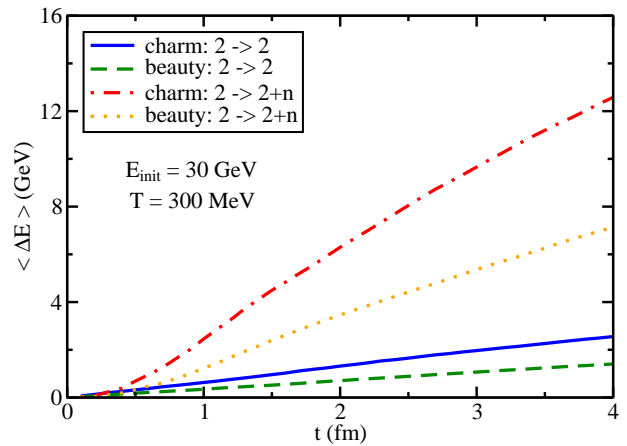}
  \caption{(Color online) Elastic vs. inelastic energy loss of heavy quarks in a static medium.}
 \label{fig:plot-tHQeLoss-real}
\end{figure}

In Fig. \ref{fig:plot-tHQeLoss-real}, we compare the energy loss of heavy quarks from $2\rightarrow2$ (collisional) and $2\rightarrow2+n$ (radiative) processes in a static medium. The initial energy of heavy quarks is set as 30~GeV, the medium temperature is fixed at 300~MeV, and $\alpha_s=0.3$ is used. For the results presented from now, the realistic variation of the heavy quark momentum during its propagation is included and $t_i$ is also updated for each heavy quark after it emits gluons. As shown in Fig. \ref{fig:plot-tHQeLoss-real}, elastic and inelastic energy loss are comparable to each other at early time. On the other hand, for large $t$, the inelastic process dominates the heavy quark energy loss. In addition, charm quarks lose significantly more energy than beauty quarks in both elastic and inelastic processes because of the mass effect. These results are consistent with our previous calculations based on a Langevin transport framework \cite{Cao:2012au,Cao:2015hia}. We have noted that the relative contribution of elastic vs. inelastic scattering to heavy quark energy loss may depend on the kinematic cut-off one implements. However, to better distinguish between the contribution from the two mechanisms and place more constraints on our theoretical model, new observables beyond current measurement -- such as two particle correlation functions related to heavy mesons \cite{Cao:2015cba} -- should be investigated. This will be explored in a follow-up study.

To combine elastic and inelastic processes in our model, we divide the scattering probability within each time step $\Delta t$ into two regions: pure elastic scattering with probability $P_\mathrm{el}(1-P_\mathrm{inel})$, and inelastic scattering with probability $P_\mathrm{inel}$. The total scattering probability is then their sum:
\begin{equation}
 \label{eq:totProb}
 P_\mathrm{tot}=P_\mathrm{el}+P_\mathrm{inel}-P_\mathrm{el}P_\mathrm{inel}.
\end{equation}
The Monte-Carlo method is adopted to determine whether a given heavy quark, during each $\Delta t$, scatters with the medium and whether the scattering is pure elastic or inelastic. After a channel is selected, the method discussed in either Sec. \ref{subsec:collision} or Sec. \ref{subsec:radiation} is used to update the energy and momentum of the heavy quark before it propagates to the next time step.

\section{Hadronization of heavy quarks}
\label{sec:hadronization}

In the previous section, we described the LBT model that includes heavy quark evolution inside a thermal QGP medium. In relativistic heavy-ion collisions, the temperature of the produced QGP decreases while it expands hydrodynamically. Around the transition temperature ($T_\mathrm{c}=165$~MeV in our work) both the fireball and heavy quarks hadronize into color neutral bound states. In this section, we present a hybrid model of fragmentation and coalescence mechanisms that we utilize to describe the hadronization of heavy quarks.

High momentum heavy quarks tend to fragment into heavy hadrons. On the other hand, low momentum heavy quarks prefer the combination with thermal partons to form hadrons. 
The relative probability between these two mechanisms can be determined by the Wigner function in the heavy-light quark coalescence model \cite{Oh:2009zj}. 
Using this probability, we can calculate the spectra of produced heavy mesons, via the heavy-light coalescence process within the coalescence model itself, and via the fragmentation process using \textsc{Pythia} simulation \cite{Sjostrand:2006za}. 
In this work, we follow our previous study \cite{Cao:2013ita,Cao:2015hia} and further improve our hybrid model of fragmentation and coalescence for heavy hadron production.

Based on the instantaneous coalescence model, the momentum spectra of produced mesons and baryons are determined by:
\begin{eqnarray}
\label{eq:recombMeson}
\frac{dN_M}{d^3p_M} \!\!&=&\!\!\! \int d^3p_1 d^3p_2 \frac{dN_1}{d^3p_1} \frac{dN_2}{d^3p_2}f^W_M(\vec{p}_1,\vec{p}_2)\delta(\vec{p}_M-\vec{p}_1-\vec{p}_2) \nonumber \\
\frac{dN_B}{d^3p_B} \!\!&=&\!\!\! \int d^3p_1 d^3p_2 d^3p_3 \frac{dN_1}{d^3p_1} \frac{dN_2}{d^3p_2} \frac{dN_3}{d^3p_3}f^W_B(\vec{p}_1,\vec{p}_2,\vec{p}_3) \nonumber \\  && \times \delta(\vec{p}_M-\vec{p}_1-\vec{p}_2-\vec{p}_3).
\end{eqnarray}
respectively, in which $dN_i/d^3p_i$ denotes the momentum distribution of the $i$-th constituent quark in the produced hadron. We assume thermal distribution for light partons on the hadronization (phase transition) hypersurface of the expanding medium. For heavy quarks, their distribution at $T_\mathrm{c}$ is taken after their in-medium evolution within the LBT model.
In Eq. (\ref{eq:recombMeson}), $f^W$ is known as the Wigner function denoting the probability for two or three particles to combine into a hadron. 
For a two-body system, the Wigner function reads
\begin{equation}
\label{eq:WignerMeson}
f_M^W(\vec{r},\vec{q})\equiv g_M \int d^3 r' e^{-i\vec{q}\cdot\vec{r}'}\phi_M(\vec{r}+\frac{\vec{r}'}{2})\phi^*_M(\vec{r}-\frac{\vec{r}'}{2}),
\end{equation}
where $g_M$ denotes the spin-color degrees of freedom of the produced meson and the variables $\vec{r}$ and $\vec{q}$ are the relative position and momentum of the two constituent quarks defined in the rest frame of the meson, i.e., the two-body center-of-mass frame.
$\phi_M$ represents the meson wavefunction, which we approximate by the ground state wavefunction of a simple harmonic oscillator, $\mathrm{exp}[-r^2/(2\sigma^2)]/(\pi \sigma^2)^{3/4}$, where the width $\sigma$ is related to the angular frequency of the oscillator $\omega$ via $\sigma= 1/\sqrt{\mu\omega}$ with $\mu= m_1m_2/(m_1+m_2)$ being the reduced mass of the two-body system. 
After averaging over the position space, we obtain the following momentum space Wigner function for the meson formation:
\begin{equation}
\label{eq:mesonWigner}
f^W_M(q^2) = g_M \frac{(2\sqrt{\pi}\sigma)^3}{V} e^{-q^2 \sigma^2}.
\end{equation}
We may straightforwardly generalize the above procedure to a three-body system for the baryon formation. After combining two quarks first and then combining their center of mass with the third quark, we obtain:
\begin{equation}
\label{eq:baryonWigner}
f^W_B(q_1^2,q_2^2) = g_B \frac{(2\sqrt{\pi})^6(\sigma_1\sigma_2)^3}{V^2} e^{-q_1^2 \sigma_1^2-q_2^2\sigma_2^2},
\end{equation}
where $\vec{q}_1$ is the relative momentum between the first two quarks and $\vec{q}_2$ is the relative momentum between the third quark and the center of mass of the first two quarks, both defined in the rest frame of the final baryon. $\sigma_i=1/\sqrt{\mu_i \omega}$ is the width parameter where $\mu_1= m_1m_2/(m_1+m_2)$ and $\mu_2= (m_1+m_2)m_3/(m_1+m_2+m_3)$. In the coalescence model, we adopt thermal mass of 300~MeV for $u$ and $d$ quarks and 475~MeV for $s$ quarks. Heavy quarks are not required to be thermal, and their bare masses of 1.27~GeV for $c$ and 4.19~GeV for $b$ quarks are used. 
The contribution from thermal gluons ($m=600$~MeV) is incorporated assuming that they first split into light quark pairs and then combine with heavy quarks to form hadrons.

One of the crucial ingredients of the coalescence model is the oscillator frequency $\omega$ in the hadron wavefunction, which can be related to the charge radii of the corresponding charged hadrons \cite{Oh:2009zj}. For mesons and baryons, the relations are given by:
\begin{align}
 \label{eq:radii}
  \left\langle r_M^2 \right\rangle_\mathrm{ch} &= \frac{3}{2\omega}\frac{1}{(m_1+m_2)(Q_1+Q_2)}\left(\frac{m_2}{m_1}Q_1+\frac{m_1}{m_2}Q_2\right),\nonumber\\
 \left\langle r_B^2 \right\rangle_\mathrm{ch} &= \frac{3}{2\omega}\frac{1}{(m_1+m_2+m_3)(Q_1+Q_2+Q_3)}\nonumber\\
 &\left(\frac{m_2+m_3}{m_1}Q_1+\frac{m_3+m_1}{m_2}Q_2+\frac{m_1+m_2}{m_3}Q_3\right),\nonumber
\end{align}
where $m_i$ and $Q_i$ are the mass and charge of each constituent quark. If one uses the root mean squared charge radii 0.43~fm of $D^+$, 0.62~fm of $B^+$ and 0.39~fm of both $\Lambda_c$ and $\Lambda_b$ \cite{Oh:2009zj,Hwang:2001th,SilvestreBrac:1996bg}, we obtain $\omega=0.33$~GeV for charm and beauty mesons, 0.43~GeV for charm baryons and 0.41~GeV for beauty baryons. For a minimal model, we ignore possible variation of the hadron radii due to the medium effect as discussed in Ref. \cite{Oh:2009zj}, and assume similar $\omega$ values for different heavy meson/baryon species.

\begin{figure}[tb]
  \epsfig{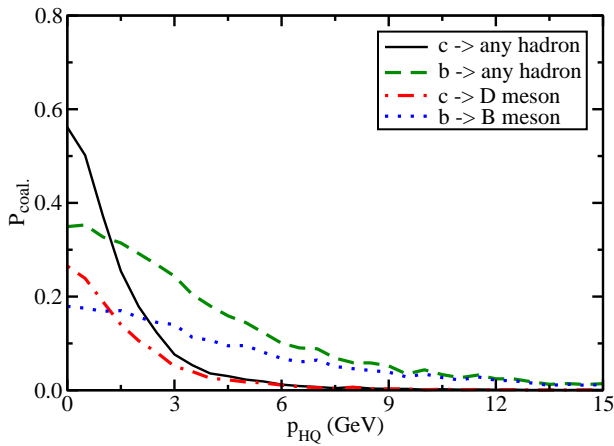}
  \caption{(Color online) The coalescence probability as a function of the heavy quark momentum.}
 \label{fig:plot-Pcoal}
\end{figure}

Using the Wigner functions in Eqs. (\ref{eq:mesonWigner}) and (\ref{eq:baryonWigner}), we can calculate the coalescence probability of heavy quarks inside a thermal medium. 
In Fig. \ref{fig:plot-Pcoal}, we provide the momentum dependence of the heavy-light quark coalescence probability at $T_\mathrm{c}$=165~MeV. 
Two sets of curves are presented: the probability for all possible hadronization channels ($D$/$B$ mesons, $\Lambda_Q$, $\Sigma_Q$, $\Xi_Q$ and $\Omega_Q$), and the one for heavy mesons ($D^0$, $D^\pm$, $B^0$, $B^\pm$) alone. 
Since $b$ quark has larger mass than $c$ quark, its coalescence probability with thermal partons is smaller than $c$ quark at zero momentum. 
On the other hand, the larger mass of $b$ quark yields a slower increase of its relative velocity with thermal partons and thus a slower decrease of its coalescence probability with respect to its momentum.

In Fig. \ref{fig:plot-Pcoal}, the hadronization of heavy quarks is divided into three regimes: coalescence with thermal partons to form $D$/$B$ mesons, coalescence to form other heavy flavor hadrons, and fragmentation. In our simulations, at $T_\mathrm{c}$, in the local rest frame of the expanding medium, we use the Monte-Carlo method to select the mechanism through which each heavy quark hadronizes. If a charm or beauty quark is selected for coalescence into a $D$ or $B$ meson, a thermal light quark or anti-quark is generated to combine with the heavy quark according to the probability governed by the Wigner function Eq. (\ref{eq:mesonWigner}). 
If they do not combine, another thermal quark is sampled until a meson is produced. 
On the other hand, if a heavy quark is selected to fragment, its conversion into heavy flavor hadron is implemented via \textsc{Pythia} fragmentation, in which the relative ratios between different hadron channels are properly calculated and normalized. 
The effects of different hadronization mechanisms on heavy meson observables in heavy-ion collisions will be investigated in detail in the next section.

\section{Heavy meson suppression and elliptic flow at the LHC and RHIC}
\label{sec:results}

In this section, we study the medium modification of heavy meson production in relativistic heavy-ion collisions. 
In particular, we investigate how heavy meson observables are affected by different momentum and temperature dependences of heavy quark transport coefficient, different hadronization mechanisms, and the coupling to the local fluid velocity of the expanding medium.

The majority of heavy quarks are produced via hard scatterings at the early stage of heavy-ion collisions. 
We initialize their production vertices in the position space via a Monte-Carlo Glauber model and their momentum distribution using the leading-order perturbative QCD (LO pQCD) calculation \cite{Combridge:1978kx}. 
Both the pair production ($gg\rightarrow Q\bar{Q}$ and $q\bar{q}\rightarrow Q\bar{Q}$) and the flavor excitation processes ($gQ\rightarrow gQ$ and $g\bar{Q}\rightarrow g\bar{Q}$) are included in our calculation of the initial $p_\mathrm{T}$ spectra of heavy quarks. 
Possible effects from the next-to-leading-order production of heavy quarks, such as the contribution of the gluon splitting process ($g\rightarrow Q\bar{Q}$), on the final state heavy meson observables have been discussed in our earlier work \cite{Cao:2015kvb} and shown to be small. 
For parton distribution functions in heavy-ion collisions, we use the CTEQ parameterizations \cite{Lai:1999wy}, together with the EPS09 parametrizations \cite{Eskola:2009uj} to include the initial state nuclear shadowing effect. 
The effect of nuclear shadowing on the initial and final heavy flavor $p_\mathrm{T}$ spectra has been presented in Ref. \cite{Cao:2015hia}. 
The rapidity distributions of initial heavy quarks are assumed to be uniform in the central rapidity region ($-1<y<1$).

To study the heavy quark evolution in realistic QGP fireballs, we couple our LBT model to the expanding bulk matter as simulated via a realistic hydrodynamic model. The (2+1)-dimensional viscous hydrodynamic model VISHNew developed in Refs. \cite{Song:2007fn,Song:2007ux,Qiu:2011hf} is adopted, unless otherwise specified. We use the code version and parameter values provided by Ref. \cite{Qiu:2011hf} in this work. The QGP fireballs are initialized using the Monte-Carlo Glauber model for their initial entropy density distribution. The starting time of the QGP evolution is set as $\tau_0=0.6$~fm and the shear-viscosity-to-entropy-density ratio ($\eta/s$=0.08) is tuned to describe the spectra of soft hadrons emitted from the QGP fireballs for both RHIC and the LHC environments. 
In this work, smooth averaged initial condition and fireball evolution for the bulk matter are used. 
Possible effects of the initial state fluctuations on heavy flavor observables have been discussed in our earlier study \cite{Cao:2014fna}. 
Before the QGP forms, heavy quarks are assumed to stream freely from their initial production vertices. 
The energy loss experienced in the short pre-equilibrium stage is expected to be small as compared to that in the much longer QGP phase.

During the QGP stage, the hydrodynamic simulation provides the spacetime evolution of the local temperature and flow velocity profiles of the QGP fireball. For each time step of the LBT simulation, we first boost each heavy quark into the local rest frame of the fluid cell through which it travels. In the rest frame of the fluid cell, the energy and momentum of a given heavy quark are updated according to the Boltzmann equation as discussed in Sec. \ref{sec:LBT}. Then the heavy quark's momentum is boosted back to the collision frame and propagates to the next time step.

In our LBT model, the only adjustable parameter is the strong coupling constant $\alpha_s$. It directly determines the cross sections and rates of elastic scatterings, and also affects the spectra of medium-induced gluon through $\hat{q}$ [Eq. (\ref{eq:22coefficients})] in the inelastic process. However, as pointed out by Refs. \cite{Gossiaux:2008jv,He:2011qa,Xu:2014tda,Das:2015ana}, the perturbative calculation of the scattering cross sections alone is not sufficient to describe heavy meson suppression and elliptic flow observed at RHIC and the LHC. To modify the temperature and momentum dependences of the pQCD driven transport coefficient in the non-perturbative region, we introduce the following Gaussian parametrizations of $K$-factors to model possible non-perturbative effect at low momentum and temperature near $T_\mathrm{c}$:
\begin{eqnarray}
 \label{eq:KP}
 K_p&=&1+A_p e^{-|\vec{p}|^2/2\sigma_p^2},\\
 \label{eq:KT}
 K_T&=&1+A_T e^{-(T-T_\mathrm{c})^2/2\sigma_T^2},
\end{eqnarray}
where the temperature dependent factor $K_T$ is directly placed on the coupling constant $\tilde{\alpha}_s=K_T\alpha_s$ and the momentum dependent factor $K_p$ is placed on the obtained transport coefficient $\tilde{\hat{q}}=K_p\hat{q}$. These $K$ factors are selected such that at high $p$ and $T$ they smoothly return to unity and thus the transport coefficient based on LO pQCD calculation is strictly applied to heavy quark evolution in the perturbative regime. The amplitude ($A$) and width ($\sigma$) parameters help model the possible strength and range of the non-perturbative effects at low momentum and temperature near $T_\mathrm{c}$. Note that a precise extraction of these parameters from the model-to-data comparison is not the purpose of this work, and will be left for a follow-up study using the Bayesian method \cite{Bernhard:2015hxa}. In this work, we only concentrate on discussing possible effects on heavy meson observables from different momentum and temperature dependences of the transport coefficient with the help of the above parametrizations.

After heavy quarks travel outside the QGP medium, i.e., the fluid cells have local temperature below $T_\mathrm{c}$, they form heavy flavor hadrons based on our hybrid model of fragmentation and coalescence as discussed in Sec. \ref{sec:hadronization}. The effects of hadronic rescatterings on heavy meson observables have been investigated in our earlier work \cite{Cao:2015hia} and shown to be small, and thus is neglected in the current calculations.
The nuclear modification factor $R_\mathrm{AA}$ and elliptic flow coefficient $v_2$ of heavy flavor mesons are calculated as follows:
\begin{align}
& R_\mathrm{AA}(p_\mathrm{T})\equiv\frac{1}{N_\mathrm{coll}}\frac{{dN^\mathrm{AA}}/{dp_\mathrm{T}}}{{dN^\mathrm{pp}}/{dp_\mathrm{T}}}, \\
& v_2(p_\mathrm{T})\equiv\langle \cos(2\phi)\rangle=\left\langle\frac{p_x^2-p_y^2}{p_x^2+p_y^2}\right\rangle.
\end{align}
Below we present our numerical results and compare to the data from the LHC and RHIC experiments.

\subsection{Effects of the momentum dependence of the transport coefficient}
\label{subsec:pDep}

\begin{figure}[tb]
 \subfigure{\label{fig:plot-RAA_LHC-0-7d5-KP}
  \epsfig{file=plot-RAA_LHC-0-7d5-KP.eps, width=0.45\textwidth, clip=}}
 \subfigure{\label{fig:plot-v2_LHC-30-50-KP}
  \epsfig{file=plot-v2_LHC-30-50-KP.eps, width=0.45\textwidth, clip=}}
  \caption{(Color online) Effects of the momentum dependence of the transport coefficient on $D$ meson (a) $R_\mathrm{AA}$ and (b) $v_2$ in Pb-Pb collisions at the LHC.}
  \label{fig:plot-LHC-KP}
\end{figure}

\begin{figure}[tb]
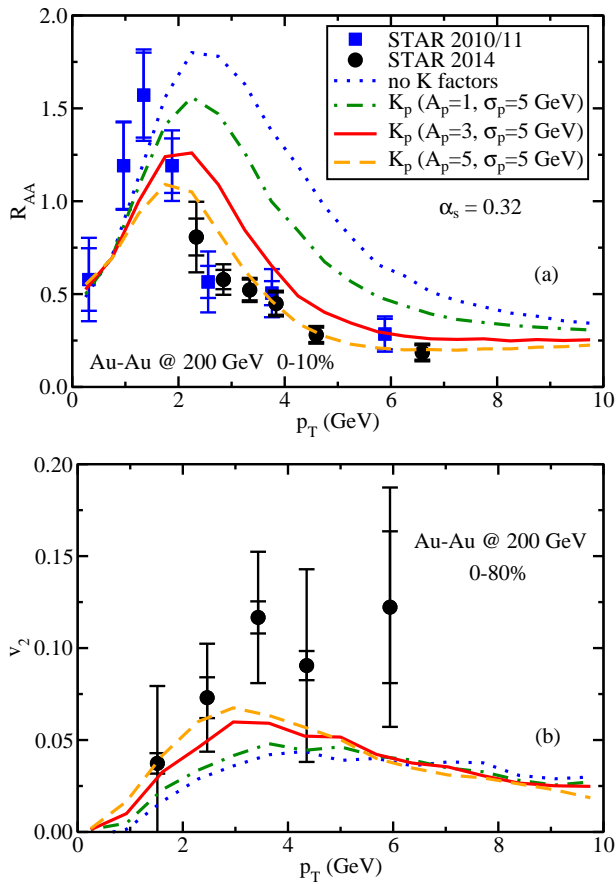

 \subfigure{\label{fig:plot-RAA_RHIC-00-10-KP}
  \epsfig{file=plot-RAA_RHIC-00-10-KP.eps, width=0.45\textwidth, clip=}}
 \subfigure{\label{fig:plot-v2_RHIC-00-80-KP}
  \epsfig{file=plot-v2_RHIC-00-80-KP.eps, width=0.45\textwidth, clip=}}
  \caption{(Color online) Effects of the momentum dependence of the transport coefficient on $D$ meson (a) $R_\mathrm{AA}$ and (b) $v_2$ in Au-Au collisions at RHIC.}
  \label{fig:plot-RHIC-KP}
\end{figure}

In Fig. \ref{fig:plot-LHC-KP}, we present our calculation of the $D$ meson $R_\mathrm{AA}$ and $v_2$ in 2.76~TeV Pb-Pb collisions at the LHC. Using the strong coupling constant $\alpha_s=0.32$, our Boltzmann model provides a good description of high $p_\mathrm{T}$ ($>$10~GeV) suppression of $D$ mesons compared to the ALICE data. 
However, at low $p_\mathrm{T}$, our model calculation underestimates both the suppression and the elliptic flow coefficient of $D$ mesons, indicating insufficient energy loss of heavy quarks when only the perturbative contribution to the transport coefficient is considered. 
We then use three sets of the momentum dependent $K_p$ factor to modify the heavy quark transport coefficient according to Eq. (\ref{eq:KP}). 
One can see that, with an enhancement of $\hat{q}$ at low momentum, better agreements of the $D$ meson $R_\mathrm{AA}$ and $v_2$ with the experimental observations may be obtained. 
However, even with very large $K_p$ ($A_p=5$ and $\sigma_p=5$~GeV), $v_2$ of $D$ mesons is still underestimated while the suppression has already been slightly overestimated. 
This suggests that the use of the $K_p$ factor alone is not able to describe the complete non-perturbative effect for the heavy quark transport coefficient.
Similar observations can also be found in Fig. \ref{fig:plot-RHIC-KP} where the $D$ meson suppression and elliptic flow are shown for 200~GeV Au-Au collisions at RHIC:
the use of $\hat{q}$ of heavy quarks obtained from pQCD calculation is inadequate to describe the $D$ meson data; an enhancement of $\hat{q}$ at low momentum can provide a reasonable description of the $D$ meson $R_\mathrm{AA}$, but a hint of underestimation of the $D$ meson $v_2$ still exists.

\subsection{Effects of the temperature dependence of the transport coefficient}
\label{subsec:TDep}

\begin{figure}[tb]
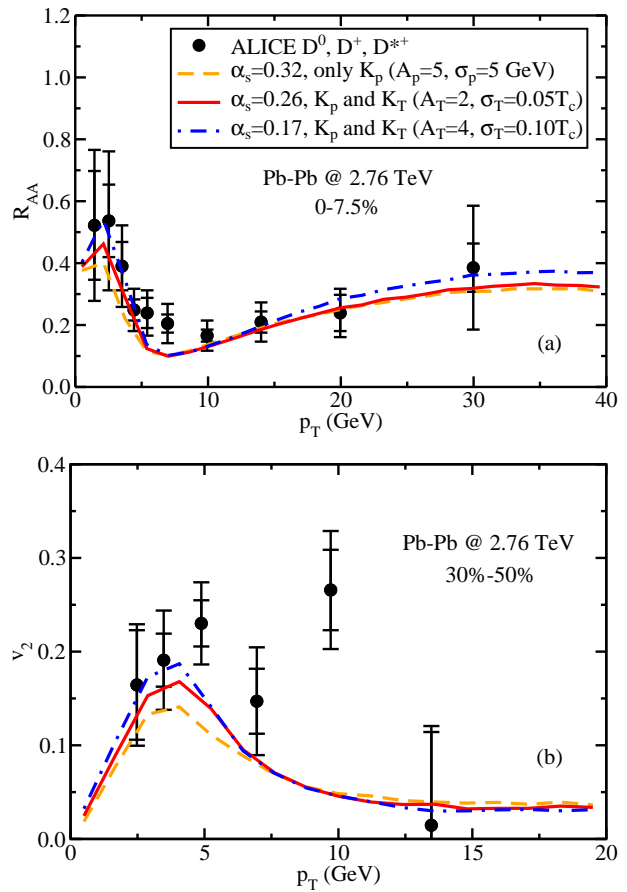

 \subfigure{\label{fig:plot-RAA_LHC-0-7d5-KT}
  \epsfig{file=plot-RAA_LHC-0-7d5-KT.eps, width=0.45\textwidth, clip=}}
 \subfigure{\label{fig:plot-v2_LHC-30-50-KT}
  \epsfig{file=plot-v2_LHC-30-50-KT.eps, width=0.45\textwidth, clip=}}
  \caption{(Color online) Effects of the temperature dependence of the transport coefficient on $D$ meson (a) $R_\mathrm{AA}$ and (b) $v_2$ in Pb-Pb collisions at the LHC.}
  \label{fig:plot-LHC-KT}
\end{figure}

\begin{figure}[tb]
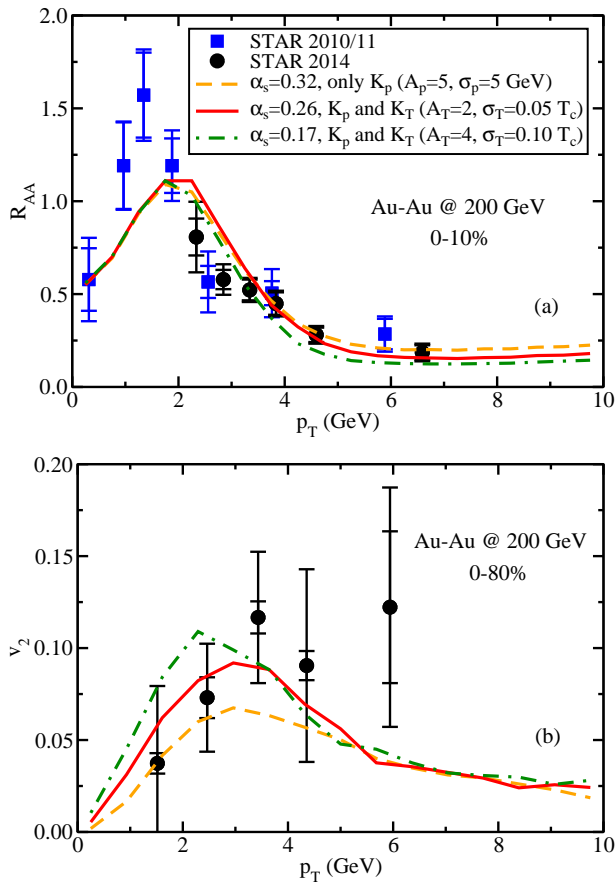

 \subfigure{\label{fig:plot-RAA_RHIC-00-10-KT}
  \epsfig{file=plot-RAA_RHIC-00-10-KT.eps, width=0.45\textwidth, clip=}}
 \subfigure{\label{fig:plot-v2_RHIC-00-80-KT}
  \epsfig{file=plot-v2_RHIC-00-80-KT.eps, width=0.45\textwidth, clip=}}
  \caption{(Color online) Effects of the temperature dependence of the transport coefficient on $D$ meson (a) $R_\mathrm{AA}$ and (b) $v_2$ in Au-Au collisions at RHIC.}
   \label{fig:plot-RHIC-KT}
\end{figure}

As discussed above, the use of the $K_p$ factor alone is insufficient to parametrize the full non-perturbative effect on heavy quark interaction with the QGP medium. 
We now investigate how the heavy meson observables are affected by the temperature dependent modifications of the transport coefficient. 
For this purpose, we multiply the constant $\alpha_s$ by the $K_T$ factor as defined by Eq. (\ref{eq:KP}). 
The $K_p$ factor is now fixed at $A_p=5$ and $\sigma_p=5$~GeV, and we use three sets of values for the parameters $A_T$ and $\sigma_T$.

As shown by Figs. \ref{fig:plot-RAA_LHC-0-7d5-KT} and \ref{fig:plot-RAA_RHIC-00-10-KT}, we can adjust the value of $\alpha_s$ such that different choices of $K_T$ provide similar $R_\mathrm{AA}$ of $D$ mesons at both the LHC and RHIC. However, three sets of parameters of $K_T$ produce very different $v_2$: a stronger enhancement of the transport coefficient near $T_\mathrm{c}$, i.e., larger values of $A_T$ and $\sigma_T$, yields larger $v_2$ for $D$ mesons while their $R_\mathrm{AA}$ are quite similar. 
One may understand this as follows. 
If one increases the transport coefficient around $T_\mathrm{c}$ while keeping the overall suppression of heavy mesons fixed, larger part of heavy quark energy loss tends to be shifted to the moment when they travel across the QGP boundary (or phase transition hypersurface), where the collective flow of the bulk matter is larger than average. 
Thus heavy quarks obtain larger $v_2$ due to stronger interaction around $T_\mathrm{c}$. 
Our results are qualitatively consistent with the findings in Refs. \cite{Xu:2014tda,Das:2015ana}.

\begin{figure}[tb]
  \epsfig{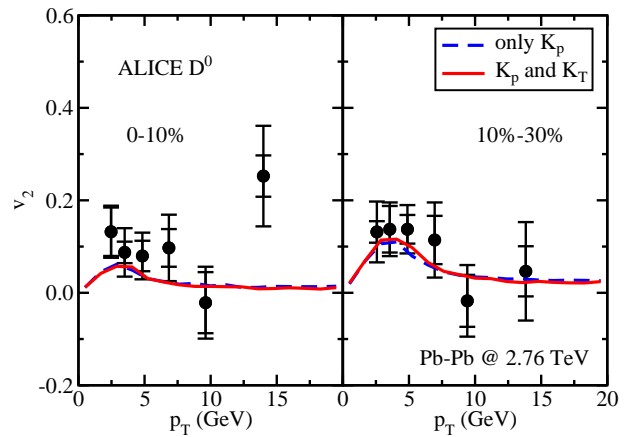}
  \caption{(Color online) $D$ meson $v_2$ in 0-10\% and 10\%-30\% Pb-Pb collisions at the LHC.}
 \label{fig:plot-v2_LHC-2cen}
\end{figure}

\begin{figure}[tb]
  \epsfig{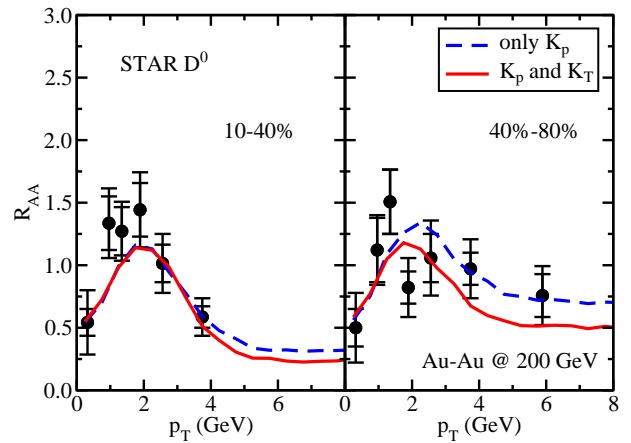}
  \caption{(Color online) $D$ meson $R_\mathrm{AA}$ in 10-40\% and 40\%-80\% Au-Au collisions at RHIC.}
 \label{fig:plot-RAA_RHIC-2cen}
\end{figure}

As shown by Figs. \ref{fig:plot-LHC-KT} and \ref{fig:plot-RHIC-KT}, with proper choices of both $K_p$ and $K_T$ (e.g., $A_p=5$, $\sigma_p=5$~GeV, $A_T=2$, $\sigma_T=0.05T_\mathrm{c}$), our LBT calculation can provide reasonable descriptions of the $D$ meson $R_\mathrm{AA}$ in central collisions and $v_2$ in peripheral/minimum-bias collisions at the LHC and RHIC. 
With the same parameters above\footnote{$\alpha_s=0.32$ when only $K_p$ is included, $\alpha_s=0.26$ when both $K_p$ and $K_T$ are included.}, we show in Figs. \ref{fig:plot-v2_LHC-2cen} and \ref{fig:plot-RAA_RHIC-2cen} the calculations of the $D$ meson elliptic flow and suppression in other centrality bins at the LHC and RHIC. 
One can see that our results agree with the experimental data quite well. 
Note that apart from the possible increase of the $D$ meson $v_2$ with the inclusion of $K_T$, another effect is the reduction of the $D$ meson $R_\mathrm{AA}$ in peripheral collisions while its value in central collisions is fixed. 
Therefore a global fit to experimental data would be necessary to precisely extract the non-perturbative effect. We will leave this to a future study.

\subsection{Effects of different hadronization processes}
\label{subsec:fragVSrecomb}

\begin{figure}[tb]
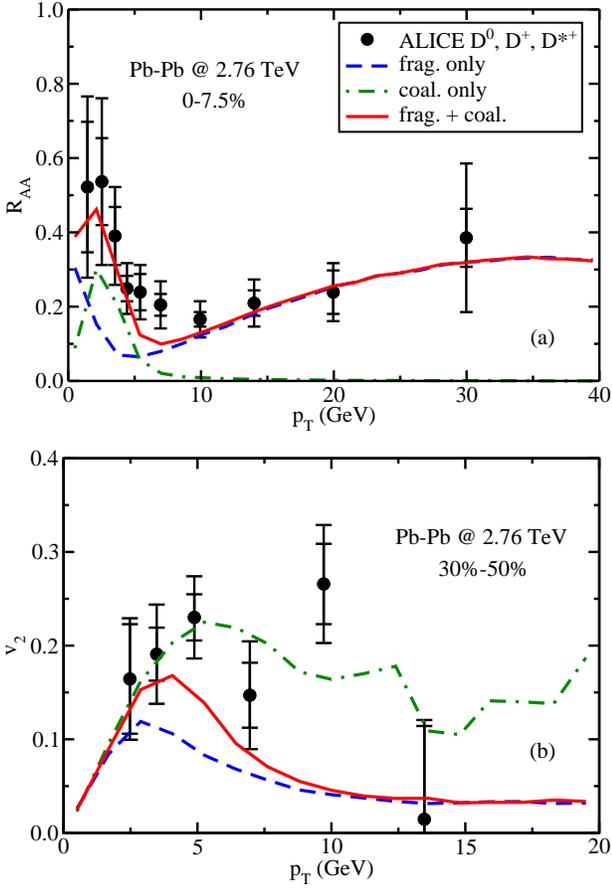

 \subfigure{\label{fig:plot-RAA_LHC-0-7d5-hadr}
  \epsfig{file=plot-RAA_LHC-0-7d5-hadr.eps, width=0.45\textwidth, clip=}}
 \subfigure{\label{fig:plot-v2_LHC-30-50-hadr}
  \epsfig{file=plot-v2_LHC-30-50-hadr.eps, width=0.45\textwidth, clip=}}
  \caption{(Color online) Effects of fragmentation vs. coalescence process on $D$ meson (a) $R_\mathrm{AA}$ and (b) $v_2$ in Pb-Pb collisions at the LHC.}
  \label{fig:plot-LHC-hadr}
\end{figure}

\begin{figure}[tb]
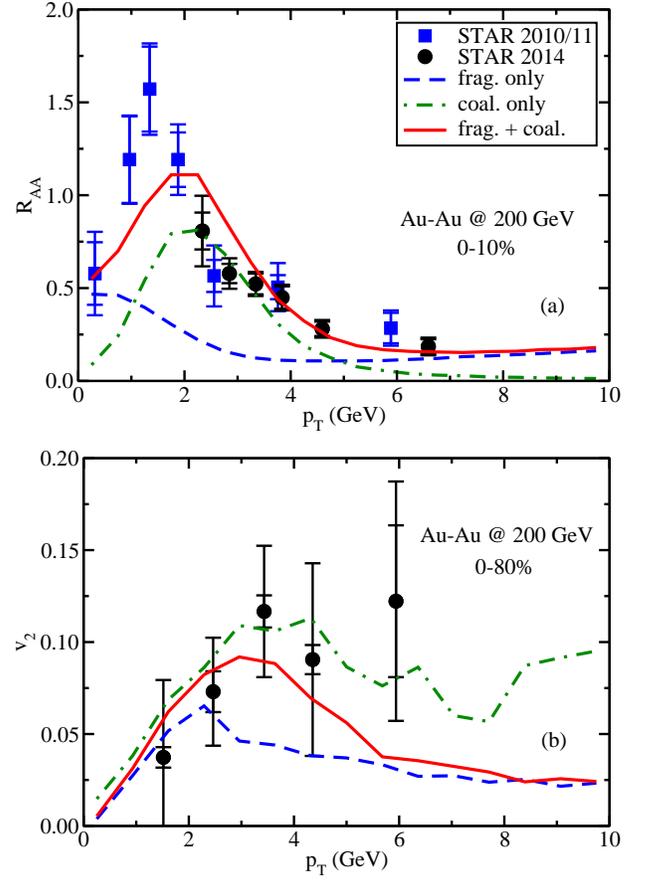

 \subfigure{\label{fig:plot-RAA_RHIC-00-10-hadr}
  \epsfig{file=plot-RAA_RHIC-00-10-hadr.eps, width=0.45\textwidth, clip=}}
 \subfigure{\label{fig:plot-v2_RHIC-00-80-hadr}
  \epsfig{file=plot-v2_RHIC-00-80-hadr.eps, width=0.45\textwidth, clip=}}
  \caption{(Color online) Effects of fragmentation vs. coalescence process on $D$ meson (a) $R_\mathrm{AA}$ and (b) $v_2$ in Au-Au collisions at RHIC.}
   \label{fig:plot-RHIC-hadr}
\end{figure}

Now we investigate the effects of different hadronization mechanisms on the final $D$ meson observables. 
For the remaining results presented in this work, the transport coefficient of heavy quark is calculated with $\alpha_s=0.26$, $A_p=5$, $\sigma_p=5$~GeV, $A_T=2$ and $\sigma_T=0.05T_\mathrm{c}$.

From Figs. \ref{fig:plot-RAA_LHC-0-7d5-hadr} and \ref{fig:plot-RAA_RHIC-00-10-hadr}, we observe that the fragmentation mechanism dominates the $D$ meson production at high $p_\mathrm{T}$ ($>8$~GeV) at both the LHC and RHIC. On the other hand, the coalescence mechanism tends to combine low momentum charm quarks and thermal partons, thus significantly enhances the $D$ meson yield in the intermediate $p_\mathrm{T}$ region ($1\sim4$~GeV). 
For this reason, one may observe the bump structure of the $D$ meson $R_\mathrm{AA}$ which is hard to obtain with  fragmentation alone. In addition, as shown by Figs. \ref{fig:plot-v2_LHC-30-50-hadr} and \ref{fig:plot-v2_RHIC-00-80-hadr}, the heavy-light coalescence mechanism can produce a much larger $D$ meson $v_2$ than the fragmentation mechanism, since the coalescence adds the momentum space anisotropy of light partons onto heavy quarks during hadronization, thus enhances $v_2$ of the produced $D$ mesons.

\subsection{Effects of the radial flow of the medium}
\label{subsec:flow}

\begin{figure}[tb]
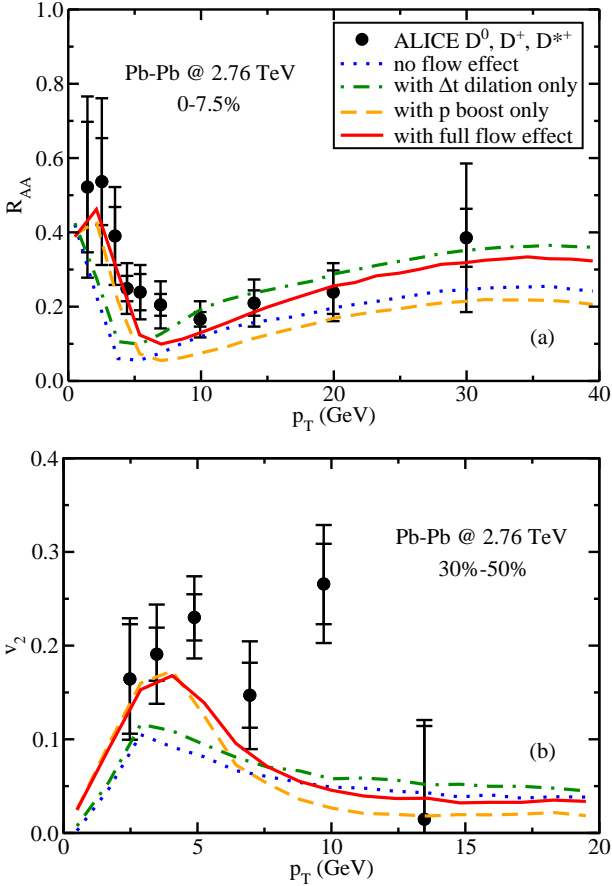

 \subfigure{\label{fig:plot-RAA_LHC-0-7d5-flow}
  \epsfig{file=plot-RAA_LHC-0-7d5-flow.eps, width=0.45\textwidth, clip=}}
 \subfigure{\label{fig:plot-v2_LHC-30-50-flow}
  \epsfig{file=plot-v2_LHC-30-50-flow.eps, width=0.45\textwidth, clip=}}
  \caption{(Color online) Effects of the medium flow on $D$ meson (a) $R_\mathrm{AA}$ and (b) $v_2$ in Pb-Pb collisions at the LHC.}
  \label{fig:plot-LHC-flow}
\end{figure}

\begin{figure}[tb]
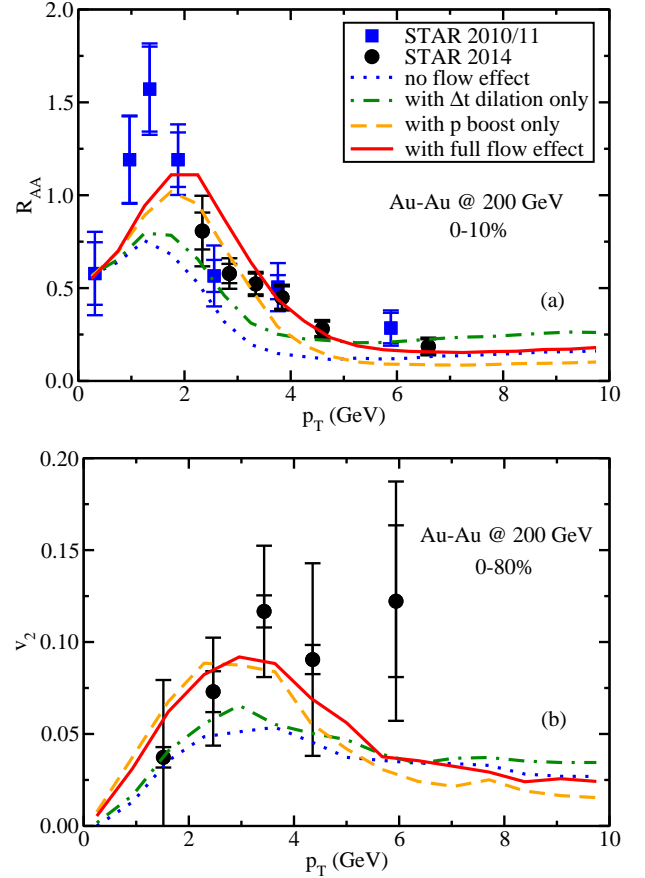

 \subfigure{\label{fig:plot-RAA_RHIC-00-10-flow}
  \epsfig{file=plot-RAA_RHIC-00-10-flow.eps, width=0.45\textwidth, clip=}}
 \subfigure{\label{fig:plot-v2_RHIC-00-80-flow}
  \epsfig{file=plot-v2_RHIC-00-80-flow.eps, width=0.45\textwidth, clip=}}
  \caption{(Color online) Effects of the medium flow on $D$ meson (a) $R_\mathrm{AA}$ and (b) $v_2$ in Au-Au collisions at RHIC.}
  \label{fig:plot-RHIC-flow}
\end{figure}

In this subsection, we investigate in detail how the radial flow of the expanding QGP medium affects the $D$ meson observables. 
There are two effects to be taken into account for the scattering and radiation rates when the transport model is coupled to a dynamically evolving medium. 
First, the time interval $\Delta t'$ in the local rest frame of the fluid cell is different from $\Delta t$ in the global collision frame by a factor of $\Delta t'/\Delta t=p_\mu u^\mu/p_0$, where $p^\mu$ is the 4-momentum of the parton under discussion and $u^\mu$ is the 4-velocity of the fluid cell in the global frame.
We denote this first effect as ``$\Delta t$ dilation". 
Second, the 4-momentum of the parton in the global frame should be boosted into the local rest frame first in which its energy and momentum are updated before it is boosted back into the global frame. 
This second effect is denoted as ``$p$ boost".
One alternative way of incorporating the ``$\Delta t$ dilation" effect is to calculate the parton energy loss still in the global frame with $\Delta t$ but rescaling the transport coefficient by $\hat{q}_\mathrm{global}/\hat{q}_\mathrm{local}=p^\mu u_\mu/p_0$, since the transverse momentum broadening of a parton $\hat{q}\Delta t$ is boost invariant \cite{Baier:2006pt}. 
For the ``$p$ boost" effect, it is either automatically included if the transport model has been organized in a boost-invariant form, otherwise it must be treated separately (e.g. the expression for the gluon differential radiation rate governing the energy loss of hard parton in Eq. (\ref{eq:gluondistribution}) is not boost-invariant).

One may compare the solutions of the transport equation using the global frame and the local rest frame \cite{Cao:2012jt}, and study how the radial flow of the expanding medium affects the final heavy meson $R_\mathrm{AA}$ and $v_2$ as shown in Figs. \ref{fig:plot-LHC-flow} and \ref{fig:plot-RHIC-flow}. 
For the $D$ meson $R_\mathrm{AA}$, we observe that due to ``$\Delta t$ dilation", heavy quarks have less time losing energy inside an expanding medium and therefore it increases as compared to the case without the flow effect. 
On the other hand, ``$p$ boost" on average enhances heavy quark energy loss in our model and therefore decreases the $D$ meson $R_\mathrm{AA}$ at high $p_\mathrm{T}$. 
The combination of these two effects increases the $D$ meson $R_\mathrm{AA}$ in our model. 
We note that how these two effects compete with each other depends on the details in the energy loss formalism, especially the momentum dependence; different models may lead to slightly different results \cite{Qin:2007zzf, Renk:2005ta}.
For the $D$ meson $v_2$, since heavy quarks traveling in the $x$ direction experience stronger decrease of energy loss due to the ``$\Delta t$ dilation" effect than those in the $y$ direction, the first effect enhances the final $D$ meson $v_2$. 
On the other hand, the second effect decreases the $D$ meson $v_2$ at high $p_\mathrm{T}$. 
At low $p_\mathrm{T}$, the boost of heavy quark momentum into and out of the fluid cell develops the coupling between the heavy quark motion and the local fluid velocity, and thus significantly enhances the elliptic flow $v_2$ of heavy quarks. 
Finally, we observe that the combination of two effects enhances the $D$ meson $v_2$ at low $p_\mathrm{T}$, but produces a mild decrease of it at high $p_\mathrm{T}$.

\subsection{Predictions for Pb-Pb collisions at $\sqrt{s_\mathrm{NN}}$=5.02~TeV}
\label{subsec:prediction}

\begin{figure}[tb]
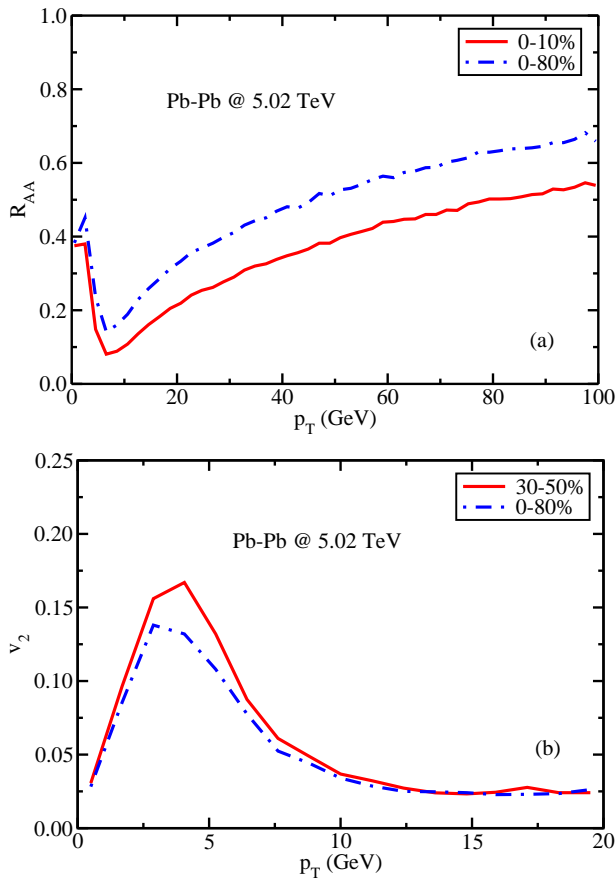

 \subfigure{\label{fig:plot-RAA_PbPb5020}
  \epsfig{file=plot-RAA_PbPb5020.eps, width=0.45\textwidth, clip=}}
 \subfigure{\label{fig:plot-v2_PbPb5020}
  \epsfig{file=plot-v2_PbPb5020.eps, width=0.45\textwidth, clip=}}
  \caption{(Color online) Predictions of $D$ meson (a) $R_\mathrm{AA}$ and (b) $v_2$ in Pb-Pb collisions at $\sqrt{s_\mathrm{NN}}=5.02$~TeV.}
  \label{fig:plot-PbPb5020}
\end{figure}

Using the same setting as described above, we now provide the predictions of the $D$ meson suppression and elliptic flow in 5.02~TeV Pb-Pb collisions in Fig. \ref{fig:plot-PbPb5020}: $R_\mathrm{AA}$ in central and minimum bias collisions are presented in Fig. \ref{fig:plot-RAA_PbPb5020} and $v_2$ in peripheral and minimum bias collisions are presented in Fig. \ref{fig:plot-v2_PbPb5020}. Here a (3+1)-dimensional viscous hydrodynamic model CLVisc \cite{Pang:2012he,Pang:2014ipa} is used to predict the spacetime evolution of the QGP fireballs in which $\tau_0=0.6$~fm, $\eta/s=0.08$ and $T_\mathrm{c}=165$~MeV are employed. In the same centrality bin, we observe the $D$ meson $R_\mathrm{AA}$ is slightly smaller and its $v_2$ is slightly larger in $\sqrt{s_\mathrm{NN}}=5.02$~GeV Pb-Pb collisions than in $\sqrt{s_\mathrm{NN}}=2.76$~GeV Pb-Pb collisions.

\section{Summary and outlook}
\label{sec:summary}

We have extended the LBT model to describe the evolution of heavy quarks  and production of final heavy flavor mesons in relativistic heavy-ion collisions. All $2\rightarrow 2$ channels are incorporated for the elastic scattering process; the medium-induced gluon radiation based on the higher-twist energy loss formalism is introduced to describe the $2\rightarrow 2+n$ inelastic process experienced by hard partons. The transport model is coupled to the hydrodynamic model that provides the spacetime evolution of QGP fireballs created in high energy nuclear collisions. A hybrid model of fragmentation and coalescence has been developed to describe the hadronization process of heavy quarks on the hadronization (or phase transition) hypersurface.

Within this LBT framework, we have investigated how heavy flavor observables depend on different energy loss and hadronization mechanisms, different momentum and temperature dependences of the jet transport coefficient, and the radial flow of QGP fireballs. 
Our results show that collisional and radiative energy losses of heavy quarks are comparable at very early time, but at later time, the evolution is more dominated by the inelastic process. 
It was also found that while the transport coefficient as calculated from the leading-order pQCD is able to describe the suppression of $D$ mesons at high $p_\mathrm{T}$ quite well, it fails at low $p_\mathrm{T}$. 
A good $p_\mathrm{T}$ dependence of the $D$ meson $R_\mathrm{AA}$ at both high and low $p_\mathrm{T}$ can be obtained by adding an enhancement for the transport coefficient at low momentum parametrized by a $K_p$ factor. 
Furthermore, a simultaneous description of the $D$ meson $R_\mathrm{AA}$ and $v_2$ requires the introduction of a temperature dependent modification ($K_T$ factor) for the transport coefficient that enhances the relative contribution of medium modification near $T_\mathrm{c}$. 
Different hadronization mechanisms also affect the final $D$ meson spectra: the fragmentation process dominates the $D$ meson production at high $p_\mathrm{T}$, while the heavy-light quark coalescence process enhances the $D$ meson yield and $v_2$ at medium $p_\mathrm{T}$. 
In addition, we have systematically analyzed the effects of the QGP radial flow on the $D$ meson observables. 
Our study indicates that the time dilation effect reduces the energy loss of heavy quarks in a moving fluid cell, and thus increases both $R_\mathrm{AA}$ and $v_2$ of $D$ mesons. 
On the other hand, the momentum boost of heavy quarks into and out of the local rest frame of the fluid cell enhances heavy quark energy loss and therefore decreases the $D$ meson $R_\mathrm{AA}$ and $v_2$ at large $p_\mathrm{T}$. 
We have found that while the combination of the two flow effects can usually enhance the $D$ meson $v_2$ at low to medium $p_\mathrm{T}$, the influence at high $p_\mathrm{T}$ is model-dependent and might be small. 
With a proper modification of the pQCD-driven transport coefficient to include some possible non-perturbative effects, our model can provide good descriptions of the $D$ meson suppression and elliptic flow coefficient simultaneously observed at the LHC and RHIC. 
We have also provided the predictions for the $D$ meson $R_\mathrm{AA}$ and $v_2$ in 5.02~TeV Pb-Pb collisions at the LHC.

The current study constitutes an important contribution to our quantitative understanding of the production and nuclear modification of heavy flavors in relativistic heavy-ion collisions. 
There are two directions to explore in the near future. 
As mentioned earlier, a systematic global fit of the experimental data is necessary in order to precisely extract the non-perturbative effect of the parton-medium interaction at low momentum and near the transition temperature $T_\mathrm{c}$. 
For this purpose, a Bayesian analysis may be applied. 
Furthermore, our LBT model has been established not only for heavy quark evolution, but also for light flavor partons. 
In an upcoming work, the nuclear modification of high energy light flavor hadrons and heavy-light hadron correlations will be explored in the same framework.

\section*{Acknowledgments}

We are grateful to Long-Gang Pang for providing the hydrodynamic profiles of Pb-Pb collisions at $\sqrt{s_\mathrm{NN}}=5.02$~TeV and to Jun Tao for valuable discussions. This work is funded by the Director, Office of Energy Research, Office of High Energy and Nuclear Physics, Division of Nuclear Physics, of the U.S. Department of Energy under Contract No. DE-AC02-05CH11231 within the framework of the JET Collaboration; by the Natural Science Foundation of China (NSFC) under Grants No. 11221504 and No. 11375072; by the Chinese Ministry of Science and Technology under Grant No. 2014DFG02050; and by the Major State Basic Research Development Program in China No. 2014CB845404.

\bibliographystyle{h-physrev5}
\bibliography{SCrefs}

\begin{thebibliography}{10}

\bibitem{Adams:2003kv}
STAR, J.~Adams {\em et~al.},
\newblock Phys. Rev. Lett. {\bf 91}, 172302 (2003), arXiv:nucl-ex/0305015.

\bibitem{Abelev:2008ae}
STAR, B.~I. Abelev {\em et~al.},
\newblock Phys. Rev. {\bf C77}, 054901 (2008), arXiv:0801.3466.

\bibitem{Adare:2012wg}
PHENIX, A.~Adare {\em et~al.},
\newblock Phys. Rev. {\bf C87}, 034911 (2013), arXiv:1208.2254.

\bibitem{Adare:2014bga}
PHENIX, A.~Adare {\em et~al.},
\newblock Phys. Rev. {\bf C92}, 034913 (2015), arXiv:1412.1043.

\bibitem{Aamodt:2010jd}
ALICE Collaboration, K.~Aamodt {\em et~al.},
\newblock Phys. Lett. {\bf B696}, 30 (2011), arXiv:1012.1004.

\bibitem{Abelev:2012di}
ALICE, B.~Abelev {\em et~al.},
\newblock Phys. Lett. {\bf B719}, 18 (2013), arXiv:1205.5761.

\bibitem{CMS:2012aa}
CMS, S.~Chatrchyan {\em et~al.},
\newblock Eur. Phys. J. {\bf C72}, 1945 (2012), arXiv:1202.2554.

\bibitem{Chatrchyan:2012xq}
CMS, S.~Chatrchyan {\em et~al.},
\newblock Phys. Rev. Lett. {\bf 109}, 022301 (2012), arXiv:1204.1850.

\bibitem{ATLAS:2011ah}
ATLAS Collaboration, G.~Aad {\em et~al.},
\newblock Phys. Lett. {\bf B707}, 330 (2012), arXiv:1108.6018,
\newblock Long author list - awaiting processing.

\bibitem{Qin:2015srf}
G.-Y. Qin and X.-N. Wang,
\newblock Int. J. Mod. Phys. {\bf E24}, 1530014 (2015), arXiv:1511.00790.

\bibitem{Wang:1991xy}
X.-N. Wang and M.~Gyulassy,
\newblock Phys. Rev. Lett. {\bf 68}, 1480 (1992).

\bibitem{Gyulassy:1993hr}
M.~Gyulassy and X.-N. Wang,
\newblock Nucl. Phys. {\bf B420}, 583 (1994), nucl-th/9306003.

\bibitem{Baier:1996kr}
R.~Baier, Y.~L. Dokshitzer, A.~H. Mueller, S.~Peigne, and D.~Schiff,
\newblock Nucl. Phys. {\bf B483}, 291 (1997), hep-ph/9607355.

\bibitem{Zakharov:1996fv}
B.~G. Zakharov,
\newblock JETP Lett. {\bf 63}, 952 (1996), arXiv:hep-ph/9607440.

\bibitem{Braaten:1991we}
E.~Braaten and M.~H. Thoma,
\newblock Phys. Rev. {\bf D44}, 2625 (1991).

\bibitem{Qin:2007rn}
G.-Y. Qin {\em et~al.},
\newblock Phys. Rev. Lett. {\bf 100}, 072301 (2008), arXiv:0710.0605.

\bibitem{Bass:2008rv}
S.~A. Bass {\em et~al.},
\newblock Phys. Rev. {\bf C79}, 024901 (2009), arXiv:0808.0908.

\bibitem{Armesto:2009zi}
N.~Armesto, M.~Cacciari, T.~Hirano, J.~L. Nagle, and C.~A. Salgado,
\newblock J. Phys. {\bf G37}, 025104 (2010), arXiv:0907.0667.

\bibitem{Chen:2010te}
X.-F. Chen, C.~Greiner, E.~Wang, X.-N. Wang, and Z.~Xu,
\newblock Phys. Rev. {\bf C81}, 064908 (2010), arXiv:1002.1165.

\bibitem{Zhang:2007ja}
H.~Zhang, J.~F. Owens, E.~Wang, and X.-N. Wang,
\newblock Phys. Rev. Lett. {\bf 98}, 212301 (2007), arXiv:nucl-th/0701045.

\bibitem{Renk:2008xq}
T.~Renk,
\newblock Phys. Rev. {\bf C78}, 034904 (2008), arXiv:0803.0218.

\bibitem{Majumder:2004pt}
A.~Majumder, E.~Wang, and X.-N. Wang,
\newblock Phys. Rev. Lett. {\bf 99}, 152301 (2007), arXiv:nucl-th/0412061.

\bibitem{Renk:2006qg}
T.~Renk,
\newblock Phys. Rev. {\bf C74}, 034906 (2006), hep-ph/0607166.

\bibitem{Zhang:2009rn}
H.~Zhang, J.~Owens, E.~Wang, and X.-N. Wang,
\newblock Phys. Rev. Lett. {\bf 103}, 032302 (2009), arXiv:0902.4000.

\bibitem{Qin:2009bk}
G.-Y. Qin, J.~Ruppert, C.~Gale, S.~Jeon, and G.~D. Moore,
\newblock Phys. Rev. {\bf C80}, 054909 (2009), arXiv:0906.3280.

\bibitem{Wang:2013cia}
X.-N. Wang and Y.~Zhu,
\newblock Phys. Rev. Lett. {\bf 111}, 062301 (2013), arXiv:1302.5874.

\bibitem{Qin:2010mn}
G.-Y. Qin and B.~Muller,
\newblock Phys. Rev. Lett. {\bf 106}, 162302 (2011), arXiv:1012.5280.

\bibitem{Dai:2012am}
W.~Dai, I.~Vitev, and B.-W. Zhang,
\newblock Phys. Rev. Lett. {\bf 110}, 142001 (2013), arXiv:1207.5177.

\bibitem{Chang:2016gjp}
N.-B. Chang and G.-Y. Qin,
\newblock (2016), arXiv:1603.01920.

\bibitem{Armesto:2011ht}
N.~Armesto {\em et~al.},
\newblock Phys. Rev. {\bf C86}, 064904 (2012), arXiv:1106.1106.

\bibitem{Burke:2013yra}
JET, K.~M. Burke {\em et~al.},
\newblock Phys. Rev. {\bf C90}, 014909 (2014), arXiv:1312.5003.

\bibitem{Adare:2010de}
PHENIX Collaboration, A.~Adare {\em et~al.},
\newblock Phys. Rev. {\bf C84}, 044905 (2011), arXiv:1005.1627.

\bibitem{Adare:2014rly}
PHENIX, A.~Adare {\em et~al.},
\newblock Phys. Rev. {\bf C91}, 044907 (2015), arXiv:1405.3301.

\bibitem{Adamczyk:2014uip}
STAR Collaboration, L.~Adamczyk {\em et~al.},
\newblock Phys. Rev. Lett. {\bf 113}, 142301 (2014), arXiv:1404.6185.

\bibitem{Xie:2016iwq}
STAR, G.~Xie,
\newblock (2016), arXiv:1601.00695.

\bibitem{Lomnitz:2016rpz}
STAR, M.~R. Lomnitz,
\newblock (2016), arXiv:1601.00743.

\bibitem{ALICE:2012ab}
ALICE Collaboration, B.~Abelev {\em et~al.},
\newblock JHEP {\bf 1209}, 112 (2012), arXiv:1203.2160.

\bibitem{Abelev:2013lca}
ALICE Collaboration, B.~Abelev {\em et~al.},
\newblock Phys. Rev. Lett. {\bf 111}, 102301 (2013), arXiv:1305.2707.

\bibitem{Dokshitzer:2001zm}
Y.~L. Dokshitzer and D.~E. Kharzeev,
\newblock Phys. Lett. {\bf B519}, 199 (2001), arXiv:hep-ph/0106202.

\bibitem{Molnar:2006ci}
D.~Molnar,
\newblock Eur. Phys. J. {\bf C49}, 181 (2007), arXiv:nucl-th/0608069.

\bibitem{Zhang:2005ni}
B.~Zhang, L.-W. Chen, and C.-M. Ko,
\newblock Phys. Rev. {\bf C72}, 024906 (2005), arXiv:nucl-th/0502056.

\bibitem{Uphoff:2011ad}
J.~Uphoff, O.~Fochler, Z.~Xu, and C.~Greiner,
\newblock Phys. Rev. {\bf C84}, 024908 (2011), arXiv:1104.2295.

\bibitem{Uphoff:2012gb}
J.~Uphoff, O.~Fochler, Z.~Xu, and C.~Greiner,
\newblock Phys. Lett. {\bf B717}, 430 (2012), arXiv:1205.4945.

\bibitem{Gossiaux:2010yx}
P.~Gossiaux, J.~Aichelin, T.~Gousset, and V.~Guiho,
\newblock J. Phys. {\bf G37}, 094019 (2010), arXiv:1001.4166.

\bibitem{Nahrgang:2013saa}
M.~Nahrgang, J.~Aichelin, P.~B. Gossiaux, and K.~Werner,
\newblock Phys. Rev. {\bf C90}, 024907 (2014), arXiv:1305.3823.

\bibitem{Svetitsky:1987gq}
B.~Svetitsky,
\newblock Phys. Rev. {\bf D37}, 2484 (1988).

\bibitem{Moore:2004tg}
G.~D. Moore and D.~Teaney,
\newblock Phys. Rev. {\bf C71}, 064904 (2005), hep-ph/0412346.

\bibitem{Akamatsu:2008ge}
Y.~Akamatsu, T.~Hatsuda, and T.~Hirano,
\newblock Phys. Rev. {\bf C79}, 054907 (2009), arXiv:0809.1499.

\bibitem{Das:2010tj}
S.~K. Das, J.-E. Alam, and P.~Mohanty,
\newblock Phys. Rev. {\bf C82}, 014908 (2010), arXiv:1003.5508.

\bibitem{He:2011qa}
M.~He, R.~J. Fries, and R.~Rapp,
\newblock Phys. Rev. {\bf C86}, 014903 (2012), arXiv:1106.6006.

\bibitem{Young:2011ug}
C.~Young, B.~Schenke, S.~Jeon, and C.~Gale,
\newblock Phys. Rev. {\bf C86}, 034905 (2012), arXiv:1111.0647.

\bibitem{Alberico:2011zy}
W.~Alberico {\em et~al.},
\newblock Eur. Phys. J. {\bf C71}, 1666 (2011), arXiv:1101.6008.

\bibitem{Cao:2013ita}
S.~Cao, G.-Y. Qin, and S.~A. Bass,
\newblock Phys. Rev. {\bf C88}, 044907 (2013), arXiv:1308.0617.

\bibitem{Cao:2015hia}
S.~Cao, G.-Y. Qin, and S.~A. Bass,
\newblock Phys. Rev. {\bf C92}, 024907 (2015), arXiv:1505.01413.

\bibitem{Song:2015ykw}
T.~Song, H.~Berrehrah, D.~Cabrera, W.~Cassing, and E.~Bratkovskaya,
\newblock Phys. Rev. {\bf C93}, 034906 (2016), arXiv:1512.00891.

\bibitem{He:2015pra}
Y.~He, T.~Luo, X.-N. Wang, and Y.~Zhu,
\newblock Phys. Rev. {\bf C91}, 054908 (2015), arXiv:1503.03313.

\bibitem{Guo:2000nz}
X.-F. Guo and X.-N. Wang,
\newblock Phys. Rev. Lett. {\bf 85}, 3591 (2000), arXiv:hep-ph/0005044.

\bibitem{Majumder:2009ge}
A.~Majumder,
\newblock Phys. Rev. {\bf D85}, 014023 (2012), arXiv:0912.2987.

\bibitem{Zhang:2003wk}
B.-W. Zhang, E.~Wang, and X.-N. Wang,
\newblock Phys. Rev. Lett. {\bf 93}, 072301 (2004), arXiv:nucl-th/0309040.

\bibitem{Auvinen:2009qm}
J.~Auvinen, K.~J. Eskola, and T.~Renk,
\newblock Phys. Rev. {\bf C82}, 024906 (2010), arXiv:0912.2265.

\bibitem{Eichten:1984eu}
E.~Eichten, I.~Hinchliffe, K.~D. Lane, and C.~Quigg,
\newblock Rev. Mod. Phys. {\bf 56}, 579 (1984),
\newblock [Addendum: Rev. Mod. Phys.58,1065(1986)].

\bibitem{Combridge:1978kx}
B.~Combridge,
\newblock Nucl. Phys. {\bf B151}, 429 (1979).

\bibitem{Wang:2009qb}
W.-T. Deng and X.-N. Wang,
\newblock Phys. Rev. {\bf C81}, 024902 (2010), arXiv:0910.3403.

\bibitem{Cao:2012au}
S.~Cao, G.-Y. Qin, S.~A. Bass, and B.~Muller,
\newblock Nucl. Phys. {\bf A904-905}, 653c (2013), arXiv:1209.5410.

\bibitem{Cao:2015cba}
S.~Cao, G.-Y. Qin, and S.~A. Bass,
\newblock Phys. Rev. {\bf C92}, 054909 (2015), arXiv:1505.01869.

\bibitem{Oh:2009zj}
Y.~Oh, C.~M. Ko, S.~H. Lee, and S.~Yasui,
\newblock Phys. Rev. {\bf C79}, 044905 (2009), arXiv:0901.1382.

\bibitem{Sjostrand:2006za}
T.~Sjostrand, S.~Mrenna, and P.~Z. Skands,
\newblock JHEP {\bf 0605}, 026 (2006), arXiv:hep-ph/0603175.

\bibitem{Hwang:2001th}
C.-W. Hwang,
\newblock Eur. Phys. J. {\bf C23}, 585 (2002), arXiv:hep-ph/0112237.

\bibitem{SilvestreBrac:1996bg}
B.~Silvestre-Brac,
\newblock Few Body Syst. {\bf 20}, 1 (1996).

\bibitem{Cao:2015kvb}
S.~Cao, G.-Y. Qin, and X.-N. Wang,
\newblock Phys. Rev. {\bf C93}, 024912 (2016), arXiv:1511.04009.

\bibitem{Lai:1999wy}
CTEQ, H.~L. Lai {\em et~al.},
\newblock Eur. Phys. J. {\bf C12}, 375 (2000), arXiv:hep-ph/9903282.

\bibitem{Eskola:2009uj}
K.~J. Eskola, H.~Paukkunen, and C.~A. Salgado,
\newblock JHEP {\bf 0904}, 065 (2009), arXiv:0902.4154.

\bibitem{Song:2007fn}
H.~Song and U.~W. Heinz,
\newblock Phys. Lett. {\bf B658}, 279 (2008), arXiv:0709.0742.

\bibitem{Song:2007ux}
H.~Song and U.~W. Heinz,
\newblock Phys. Rev. {\bf C77}, 064901 (2008), arXiv:0712.3715.

\bibitem{Qiu:2011hf}
Z.~Qiu, C.~Shen, and U.~Heinz,
\newblock Phys. Lett. {\bf B707}, 151 (2012), arXiv:1110.3033.

\bibitem{Cao:2014fna}
S.~Cao, Y.~Huang, G.-Y. Qin, and S.~A. Bass,
\newblock J. Phys. {\bf G42}, 125104 (2015), arXiv:1404.3139.

\bibitem{Gossiaux:2008jv}
P.~Gossiaux and J.~Aichelin,
\newblock Phys. Rev. {\bf C78}, 014904 (2008), arXiv:0802.2525.

\bibitem{Xu:2014tda}
J.~Xu, J.~Liao, and M.~Gyulassy,
\newblock Chin. Phys. Lett. {\bf 32}, 9 (2015), arXiv:1411.3673.

\bibitem{Das:2015ana}
S.~K. Das, F.~Scardina, S.~Plumari, and V.~Greco,
\newblock Phys. Lett. {\bf B747}, 260 (2015), arXiv:1502.03757.

\bibitem{Bernhard:2015hxa}
J.~E. Bernhard {\em et~al.},
\newblock Phys. Rev. {\bf C91}, 054910 (2015), arXiv:1502.00339.

\bibitem{Baier:2006pt}
R.~Baier, A.~H. Mueller, and D.~Schiff,
\newblock Phys. Lett. {\bf B649}, 147 (2007), arXiv:nucl-th/0612068.

\bibitem{Cao:2012jt}
S.~Cao, G.-Y. Qin, and S.~A. Bass,
\newblock J. Phys. {\bf G40}, 085103 (2013), arXiv:1205.2396.

\bibitem{Qin:2007zzf}
G.-Y. Qin {\em et~al.},
\newblock Phys. Rev. {\bf C76}, 064907 (2007), arXiv:0705.2575.

\bibitem{Renk:2005ta}
T.~Renk and J.~Ruppert,
\newblock Phys. Rev. {\bf C72}, 044901 (2005), arXiv:hep-ph/0507075.

\bibitem{Pang:2012he}
L.~Pang, Q.~Wang, and X.-N. Wang,
\newblock Phys. Rev. {\bf C86}, 024911 (2012), arXiv:1205.5019.

\bibitem{Pang:2014ipa}
L.-G. Pang, Y.~Hatta, X.-N. Wang, and B.-W. Xiao,
\newblock Phys. Rev. {\bf D91}, 074027 (2015), arXiv:1411.7767.

\end{thebibliography}

\end{document}